\documentclass[a4paper,11pt]{article}
\usepackage{pos}
\usepackage{multirow}

\title{The structure of the lightest positive-parity charmed mesons from LQCD}

\author*[a,b]{Eric B.~Gregory}
\author[c,d,e]{Feng-Kun Guo}
\author[f,g]{Christoph Hanhart}
\author[a,b,g]{Stefan Krieg}
\author[f,g]{Thomas Luu}
\affiliation[a]{J\"{u}lich Supercomputing Centre, Forschungszentrum J\"{u}lich, Wilhelm-Johnen-Stra{\ss}e, J\"{u}lich, Germany}

\affiliation[b]{Center for Advanced Simulation and Analytics (CASA), Forschungszentrum J\"{u}lich, Wilhelm-Johnen-Stra{\ss}e, J\"{u}lich, Germany}
\affiliation[c]{
CAS Key Laboratory of Theoretical Physics, Institute of Theoretical Physics, Chinese Academy of Sciences, Beijing, China}
\affiliation[d]{School of Physical Sciences, University of Chinese Academy of Sciences, Beijing, China}
\affiliation[e]{Peng Huanwu Collaborative Center for Research and Education, Beihang University, Beijing, China}
\affiliation[f]{Institute for Advanced Simulation, Forschungszentrum J\"{u}lich, J\"{u}lich, Germany}
\affiliation[g]{Helmholtz-Institut f\"{u}r Strahlen-und-Kernphysik and Bethe Center for Theoretical Physics,
Universit\"{a}t Bonn, Bonn, Germany}

\emailAdd{e.gregory@fz-juelich.de}

\abstract{
The nature of low-lying scalar and axial-vector charmed mesons has long been debated, specifically whether they are best explained as hadronic molecules or compact tetraquark systems. These two scenarios exhibit quite different features for the accessible $SU(3)$ multiplets in the scalar and axial-vector sectors.  To resolve this debate, we performed $N_f=3+1$ lattice simulations and calculated the energy 
levels of the $SU(3)$ $[6]$ and 
$[\overline{15}]$ multiplets for both the scalar and axial-vector mesons in an $SU(3)$ flavor-symmetric setting.
In both sectors we find attractive states for the [6] and repulsive interactions for the $[\overline{15}]$. This is consistent with the hadronic molecule picture, but not the compact tetraquark picture which predicts  a low-lying $[\overline{15}]$ states in the axial-vector sector but not in the scalar sector.
}

\FullConference{The 42nd International Symposium on Lattice Field Theory (LATTICE2025)\\
2-8 November 2025\\
Tata Institute of Fundamental Research, Mumbai, India\\}

\begin{document}
\maketitle

\section{Motivation}
There exist a number of experimentally-observed open-charm hadronic states that evade explanation as simple $q\overline{q}$ mesons. Specifically, the scalar $D^*_{s0}(2317)$~\cite{Aubert:2009wg}  and the axial-vector $D^*_{s1}(2460)$~\cite{Barnes:2003dj} are significantly lighter than quark model predictions~\cite{Godfrey:1985xj}.
The $D^*_0(2300)$~\cite{Abe:2003zm, Link:2003bd} and $D_1(2430)$~\cite{Abe:2003zm} are the corresponding $0^+$ and $1^+$ non-strange states. Here we would normally expect an increase of $\sim 150$ MeV when a light quark is swapped for a strange quark, but measurements show that their masses are significantly higher than this. Together these observations suggest that these states are not pure $q\overline{q}$ meson states.
The strongest candidates to explain these behaviors are four-quark exotic states configured either as compact tetraquarks,  or as hadronic molecules.

\section{Model predictions}
The question is which of these four-quark models more accurately describes these exotic states --- tetraquarks, where the tighter binding is between the two quarks and the two antiquarks, or hadronic molecules which are essentially a coupling of two mesons.

There is evidence \cite{PhysRevD.110.030001} that the $D_0^*(2300)$ actually consists of two poles, at 2105 MeV and 2451 MeV, respectively.
These four-quark models allow us to interpret the lighter of these as part of a $SU(3)$-flavor anti-triplet $[\bar 3]$, 
which, together with the $D_{s0}^*(2317)$, would give the correct mass hierarchy. The 2451 MeV pole would be part of the corresponding $SU(3)$ sextet $[6]$. Furthermore, there are theoretical suggestions \cite{Albaladejo:2016lbb}\cite{Du:2017zvv} that the $D_1(2420)$
also consists of two poles at 2247 MeV and 2555 MeV, with the lower pole again being part of a $SU(3)$ anti-triplet with the $D_{s1}(2460)$.

\subsection{Hadronic molecules}
We examine the $SU(3)$ flavor symmetric case, where the $u$,$d$, and $s$ quarks are degenerate in mass and the charm quark is heavier.
In this case a hadronic molecule is a $c\overline{q}$ meson scattering off a $q\overline{q}$ state. We get the decomposition 
\begin{equation}\label{eq:hm_decom}
[\bar{3}]\otimes \left([3]\otimes[\bar{3}]\right)=[\bar{3}]\otimes \left([{ 8}]\oplus[{\color{red}{1}}]\right)
=[{\overline{15}}]\oplus[{6}]\oplus[{\bar{3}}]\oplus[{\color{red}\bar{3}}]\ .
\end{equation}
In this discussion we will disregard the $[\overline{3}]$ arising from the scattering of the $c\overline{q}$ off the flavor-singlet meson (red in \autoref{eq:hm_decom}), since the $\eta'$ is not a Goldstone boson in QCD, and focus on the $[\overline{15}]$, $[6]$, and the remaining $[\overline{3}]$.

Our theoretical understanding of hadronic molecules comes from unitarized chiral perturbation theory (UChPT), e.g.,\cite{Kolomeitsev:2003ac}\cite{Albaladejo:2016lbb}\cite{Hyodo:2006kg}, which predicts at the $SU(3)$ flavor-symmetric point  a strongly attractive $[\overline{3}]$, an attractive $[6]$, and  a repulsive $[\overline{15}]$. This has been confirmed in the scalar sector with a lattice calculations~\cite{Gregory:2021rgy}\cite{Yeo:2024chk}. The expectation is that the axial-vector sector shows the same pattern, up to spin-symmetry violations.

\subsection{Tetraquark model}
A  discussion of the tetraquark model prediction of the axial-vector $cq\overline{q}\overline{q}$ spectrum at the $SU(3)$-flavor point is given in \cite{Guo_2025}. On this reference we base the following arguments.

In the tetraquark picture we describe a $cq\overline{q}\overline{q}$ states as a  diquark  and anti-diquark: $\{cq\} +\{\overline{q}\overline{q}\}$. For a full discussion, see, for example \cite{Dmitrasinovic:2005gc}.

A flavor decomposition of a single diquark is
\begin{equation}
{qq}\longrightarrow [3]\otimes [3] = [{{\overline{3}}}] \oplus {{[6]}}
\end{equation}
These are commonly referred to as the ``good diquark'' and `` bad diquark'' with 
\begin{center}
\begin {tabular}{ccc c }
$J^P$ & $C$ & $F$ &\\
\hline
$0^+$ & $\overline{3}$ &  $\overline{3}$ & {{`good diquark'}}\\
$1^+$ & $\overline{3}$ &  $6$ & {{` bad diquark'}}\\
\end{tabular}
\end{center}
From phenomenology and heavy-quark effective field theory we generally expect $\sim 150$ MeV difference between $0^+$ and $1^+$ fully-light diquarks:
\begin{equation}
    M_{\overline{q}\overline{q}}[1^+] \approx M_{\overline{q}\overline{q}}[0^+] + 150 \text{ MeV}  \text{\textit{c.f.} } M_{\Sigma_c} - M_{\Lambda_c} =167 \text{ MeV}.\label{eq:qqantidiquark}
\end{equation}
Heavy-light diquarks sit in a flavortriplet, with a similar mass difference:
\begin{equation}
    M_{cq}[1^+] \approx M_{cq}[0^+] +  150 \text{ MeV } 
\text{\textit{c.f. }} M_{D_{s1}} - M_{D^*_{s0}} =142 \text{ MeV}.\label{eq:cqdiquark}
\end{equation}



The flavor decomposition of a $\{qc\}\{\overline{q}\overline{q}\}$ state results in the same representations as in the hadronic molecule picture, but we associate them with \textit{diquarks}.

\begin{align}
[3]\otimes [\overline{3}]\otimes [\overline{3}] &=
[3] \otimes \left( {{[3]_A} }\oplus{{ [\overline{6}]_S}} \right)\nonumber \\
&= {{ [\overline{3}]_A \oplus [6]_A }} \oplus {{  [\overline{3}]_S \oplus [\overline{15}]_S}}\nonumber
\end{align}

The $ {{[3]_A} }$  and ${{ [\overline{6}]_S}} $  arising from $[\overline{3}]\otimes [\overline{3}] $
are the {{\textit{good}}} and  {\textit{bad}} (anti-) diquark, with the $A$ and $S$ subscripts denoting that they are respectively symmetric and anti-symmetric in flavor space. The $[\overline{3}]_S$ is expected to be strongly repulsive due to 't Hooft interaction linked to the $U_A(1)$ anomaly \cite{PhysRevD.70.096011}, and is neglected in the following arguments. 

\begin{table}[tbh]
    \centering

\begin{center}
\begin {tabular}{lc|c|c c c c}

Tetraquarks && $[cq]$ & $[\overline{q}\overline{q}]$\\
\hline
\multirow{2}{*}{Scalar }&\multirow{2}{*}{ $0^+$} & $0^+$ &  $0^+$ &   {{`good'$+$'good'}}&& $[\overline{3}]_A$,$[6]_A$\\
          &                                      & $1^+$ &  $1^+$ &  {{`bad'$+$'bad'}} & $\longleftarrow$ $\sim 300 $ MeV heavier & $[\overline{15}]_S$\\
\hline
\multirow{2}{*}{Axial-vector }&\multirow{2}{*}{ $1^+$} & $1^+$ &  $0^+$ & {{'bad'}} $+$ {{`good'}} &\multirow{2}{*}{ $ \Big\}$ approx. equal } & $[\overline{3}]_A$,$[6]_A$\\
                                                      & & $0^+$ &  $1^+$  &  {{`good'}}$+${{'bad'}} & & $[\overline{15}]_{S1}$\\
                                                      & & $1^+$ & $1^+$ &   {{`bad'$+$'bad'}} & &  $[\overline{15}]_{S2}$\\

\end{tabular}
\end{center}

    \caption{Required diquark content to make tetraquark $[\overline{3}]_A$,$[6]_A$, and $[\overline{15}_S]_2$ states in the scalar and axial-vector sectors. }
    \label{tab:diquark_content}
\end{table}

The key observation is that scalar and axial-vector tetraquarks have different diquark content for the same $SU(3)_f$ representations. For the scalar 
($J^P=0^+$) states we can either 
combine $0^+$ $\{cq\}$ with a $0^+$ 
$\{\overline{q}\overline{q}\}$ (two good diquarks), which give the $[\overline{3}]$ and $[6]$, or we could combine the $1^+$ $\{cq\}$ with a $1^+$ 
$\{\overline{q}\overline{q}\}$, which would give the $[\overline{15}]$. This leads us to a qualitative expectation of a $[\overline{15}]$, with two bad anti-diquarks, being much heavier than the $[\overline{3}]$ and $[6]$, which each have two good diquarks. This is summarized in \autoref{tab:diquark_content}. Note that there is also a doubly-bad $[\overline{15}]$. 

In contrast, in the $1^+$ axial-vector sector, the tetraquark description demands that the $[\overline{3}]$ and $[6]$ each have a bad $1^+$ 
$\{cq\}$ diquark and a good $0^+$ $\{\overline{q}\overline{q}\}$ anti-diquark. The $[\overline{15}]_s$ has also one good diquark and one bad anti-diquark. The expectation then is that the axial-vector $[\overline{15}]_s$ is roughly on equal footing with the $[\overline{3}]$ and $[6]$ .

\begin{figure}
    \centering
    \includegraphics[width=0.75\linewidth]{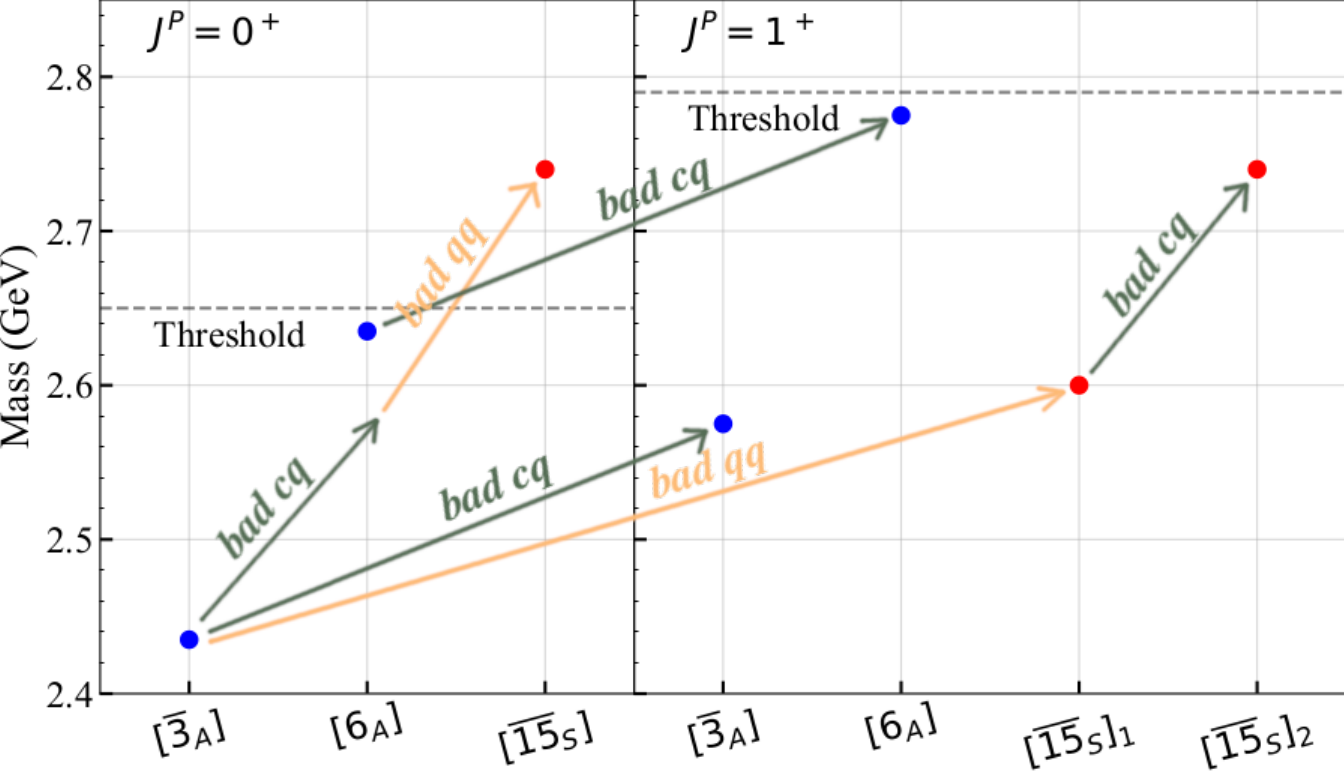}
    \caption{Plot from \cite{Guo_2025}, illustrating predicted axial-vector textraquark spectrum. Added arrow annotations show the calculated bad diquark costs. Blue (red) symbols refer to states containing spin-zero (spin-one) light
diquarks. Added green arrows show the added cost of a bad $cq$ diquark from \autoref{eq:cqdiquark}, while the yellow arrows denote the cost of adding a bad $qq$ diquark from \autoref{eq:qqantidiquark}.}
    \label{fig:guo-hanhart-prediction}
\end{figure}

In \cite{Guo_2025}, the authors begin with the spectrum of the scalar $[\overline{3}]$ and  $[6]$ at the $SU(3)$ flavor-symmetric point at $M_\pi=700$ MeV, as determined by the HadSpec collaboration \cite{Yeo:2024chk}. From these they add the phenomenologically-estimated cost of either a bad $\{qc\}$ diquark or bad $\{\overline{q}\overline{q}\}$  anti-diquark from Eqns. \ref{eq:cqdiquark} and \ref{eq:qqantidiquark}. This process is illustrated in \autoref{fig:guo-hanhart-prediction}. We see a striking difference, namely, the scalar $[\overline{15}]$ is a repulsive state, whereas  in the axial-vector sector the $[\overline{15}]$ is deeply bound and nearly degenerate with the $[\overline{3}]$.

\section{Lattice calculation}
Because of the stark differences in predictions from the two models of the energy of the axial-vector $[\overline{15}]$  state  at the $SU(3)$-flavor point, a lattice simulation is an obvious strategy to resolve the question.

As described in \cite{Gregory:2025ium}\cite{Gregory:2021rgy}\cite{Gregory:2021tjs}, we simulated $N_f=3+1$ dynamical clover ensembles with physical charm mass and three degenerate flavors of light quarks. We use six iterations of stout smearing. The calculation described here is with $\beta=3.6$ and a $64^4$ lattice. We will see that the results are informative even at a single volume and lattice spacing.

\begin{figure}
    \centering
    \includegraphics[width=0.34\linewidth]{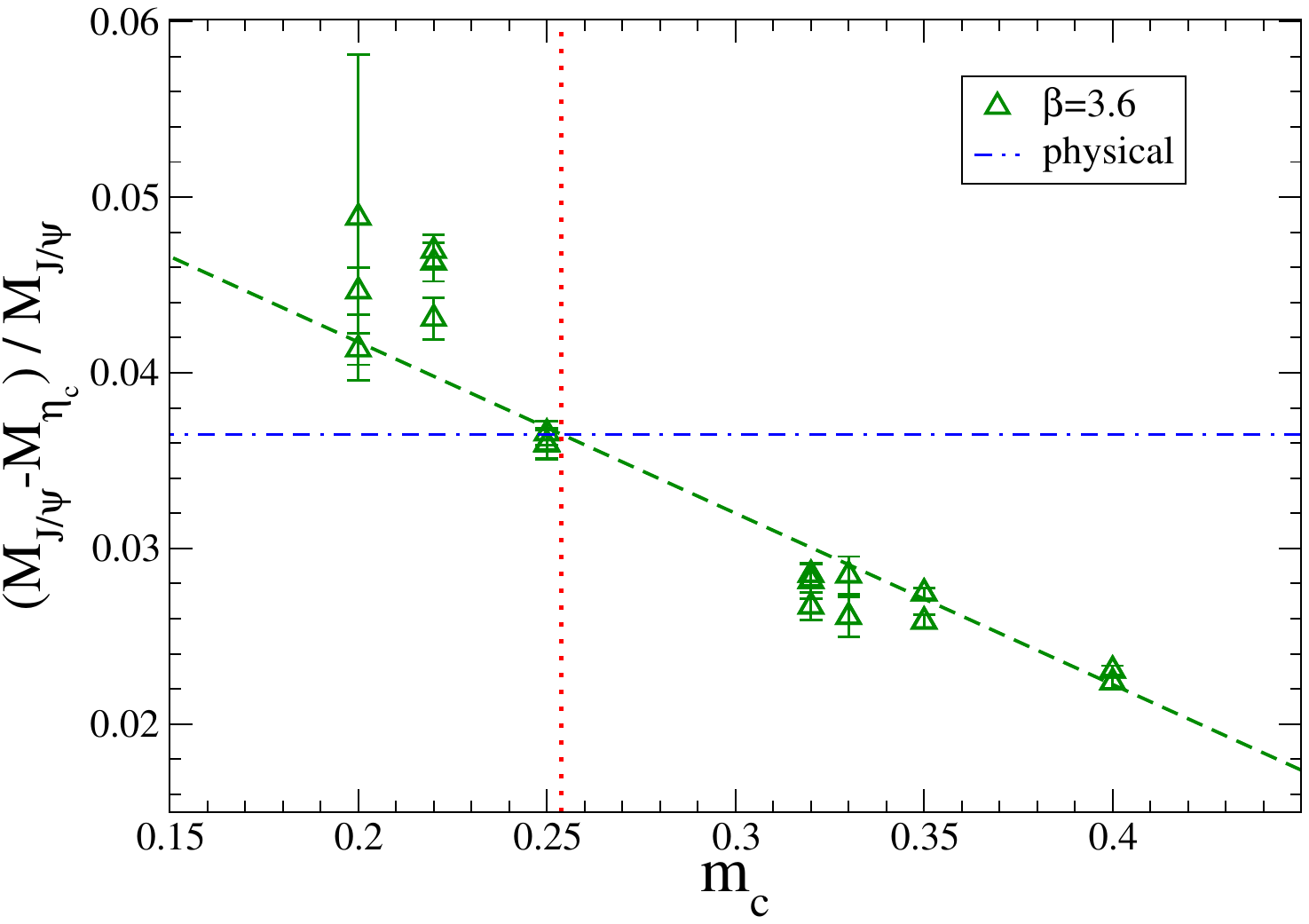}\hspace{-0.2in}
    \includegraphics[width=0.34\linewidth]{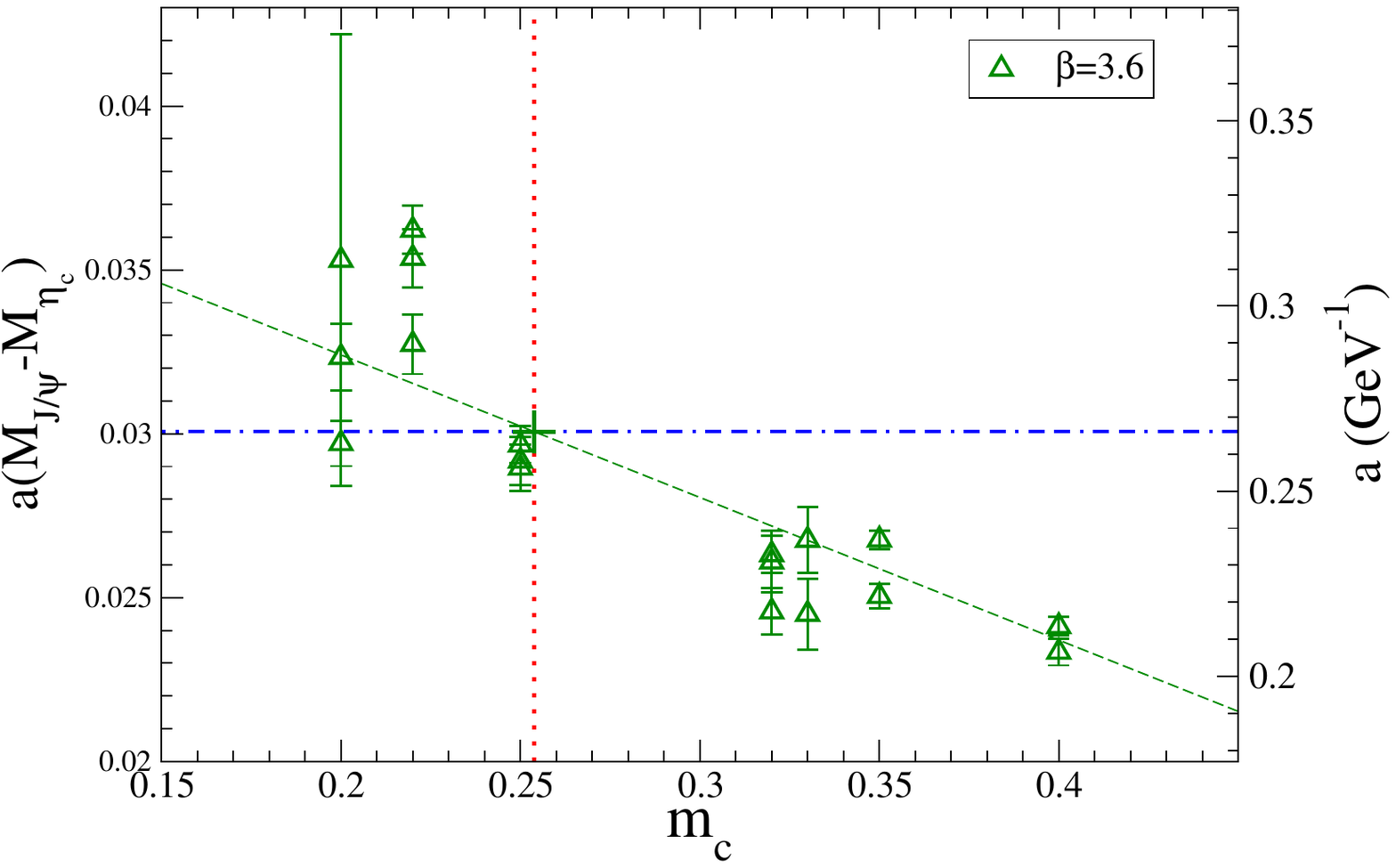}\hspace{-0.2in}
    \includegraphics[width=0.34\linewidth]{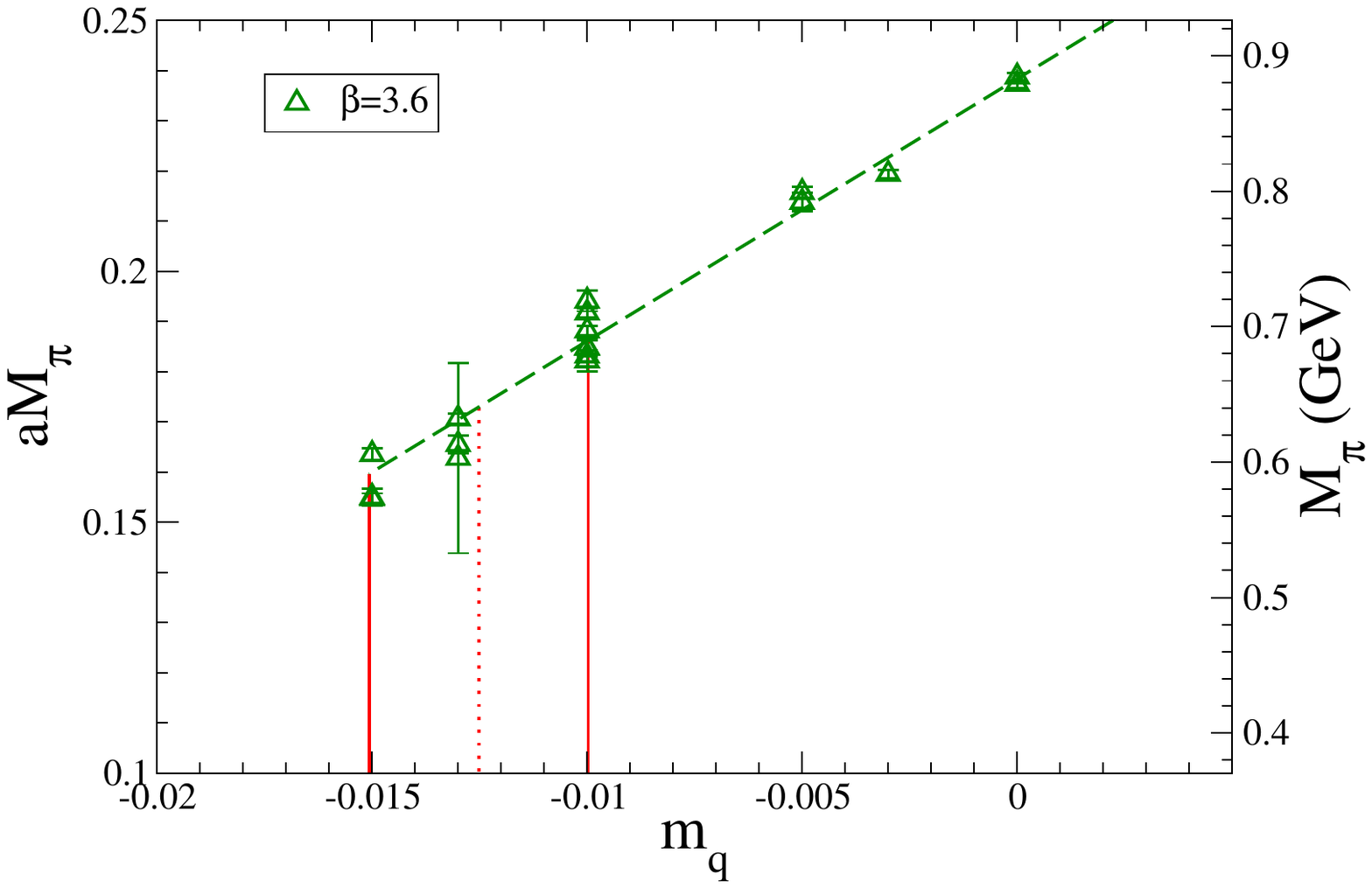}
    \caption{Tuning process for $SU(3)_f$ ensembles. First $m_c$ is tuned to give correct value of hyperfine splitting ratio (left). Hyperfine splitting ratio is used to calibrate lattice spacing $a$ (middle). Light quark is tuned to give $600 \text{ MeV } < M_{\pi}< 700$ MeV (right).)}
    \label{fig:tuning-plots}
\end{figure}

Tuning ensembles this far from physical quark light masses requires referencing charm observables. We first tuned $m_c$ until $\frac{M_{J/\psi} - M_{\eta_c}}{M_{J/\phi}}=0.0365 {\rm (phys)}$. We then extracted the lattice spacing with $a=\frac{M_{J/\psi}^{\rm latt} - M_{\eta_c}^{\rm latt}}
    {M_{J/\psi}^{\rm exp} - M_{\eta_c}^{\rm exp}}$. Finally, knowing $a$, we tuned the $m_q$ until $600 \text{ MeV } < M_{\pi}< 700$ MeV, the range at which the $[6]$ should be an attractive virtual state\cite{Du:2017zvv}. This is  also in the range of the HadSpec calculation~\cite{Yeo:2024chk}. This procedure is illustrated in \autoref{fig:tuning-plots}.

We found a target point in parameter space with  $M_\pi=613\pm 1$ MeV ($m_q=-0.013$, $m_c=0.25$) and  $a=0.27$ GeV$^{-1}$. At this target point we generated a 9960-trajectory ensemble.

\subsection{Contractions}
We performed contractions and calculated correlators for the $[6]$ and $[\overline{15}]$ in both the scalar and axial-vector sectors. We neglect the $[\overline{3}]$, which requires disconnected diagrams. The contraction for the $[6]$ is
\begin{eqnarray}
\label{eq:6contract}
\langle O^{i}_{[d]}(y';y)\bar{O}^{i}_{[d]}(x;x)\rangle^{[{{6}}]} &= \
\text{Tr}\left[\Gamma_A \gamma_5\mathcal{S}^\dag_{y';x}\gamma_5\Gamma_A\mathcal{S}^{}_{y';x}\right]
\text{Tr}\left[\Gamma_B \gamma_5\mathcal{S}^\dag_{y;x}\gamma_5\Gamma_B\mathcal{C}^{}_{y;x}\right]\\
& {{+}}\text{Tr}\left[\Gamma_B \gamma_5\mathcal{S}^\dag_{y;x}\gamma_5\Gamma_A\mathcal{S}^{}_{y';x}\Gamma_A \gamma_5\mathcal{S}^\dag_{y';x}\gamma_5\Gamma_B\mathcal{C}^{}_{y;x}\right].\nonumber
\end{eqnarray}
The contraction for the $[\overline{15}]$ is identical to \autoref{eq:6contract}, but with a relative minus sign between the terms. Here $\mathcal{S}$ is the light quark propagator, $\mathcal{C}$ is the charm propagator. In the scalar case $\Gamma_A=\Gamma_B=\gamma_5$. For the axial-vector case  $\Gamma_A= \gamma_5$, while  $\Gamma_B=\gamma_x, \gamma_y,\gamma_z$, with the resulting correlators averaged over the spatial directions $x,y,z$.

\subsection{Correlators and fits}
\begin{figure}
    \centering
    \includegraphics[width=0.8\linewidth]{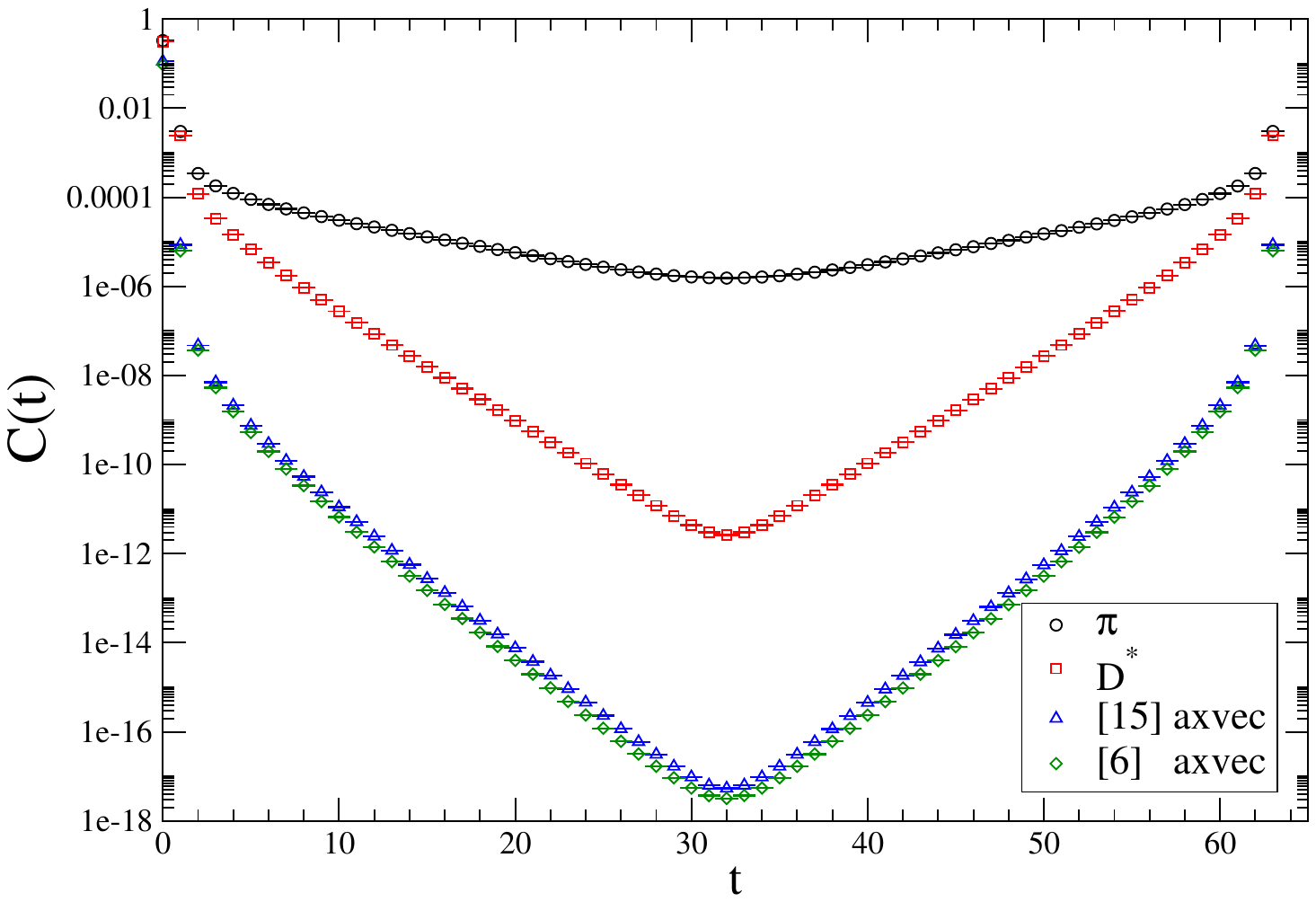}
    \caption{Representative correlators for the $\pi$, $D^*$, $[\overline{15}]_{\rm axvec}$ and $[6]_{\rm axvec}$.}
    \label{fig:ax-corrs}
\end{figure}

In addition to the  correlators for the $[6]$ and $[\overline{15}]$, we computed correlators for the $\pi$, $D$ and $D^*$ mesons. We used smeared sources and contracted with both smeared and point sinks, giving two correlators per state. After thermalization, we computed correlators on 32 sources per configuration, with a gap of 40 trajectories between measurements. \autoref{fig:ax-corrs} shows representative correlators for the $\pi$, $D^*$ and axial-vector $[6]$ and $[\overline{15}]$.

\subsection{Analysis}

For the meson states we fit the correlators to
\begin{equation}
 C_{P,s}(t)= \sum^{(N-1)}_{j=0} A_{P,s,j}\cosh\left(E_{P,j}(t-L_t/2)\right),
 \label{eq:meson-ansatz}
\end{equation}
with $P={\pi, D, D^*}$ and $s=$\{point, smeared\} sinks. We fit with a sliding  window from $t_{\rm min}\geq t \geq L_t-t_{\rm min}$, varying $t_{\rm min}$ over a range where useful fits could be obtained. For the four-quark $SU(3)$ states  we use:
\begin{equation}
      C_{P,s}(t) =
    B_{s}\cosh\left((M_{\overline{q}c}-M_\pi)(t-L_t/2)\right)
    + \sum_{j=0}^{(N-1)} A_{P,s,j}\cosh\left(E_{p,j}(t-L_t/2)\right).
\label{eq:su3-ansatz}
\end{equation}
  The $M_{\overline{q}c}$ in the first term is $M_D$ for the scalar case and $M_{D^*}$ for the axial-vector case. This term is necessary to accommodate the artifact contribution arising from the $\pi$ propagating forward in time and the $D$ or $D^*$ propagating backwards in time (or vice versa). In all cases we try fits with $N=2,3,4$ states.  
 In \autoref{eq:su3-ansatz}, $M_\pi$ and $M_{\overline{q}c}$ are fixed at values from independent fits and analysis of the corresponding meson correllators.

 For each state of interest we determine ground-state energy expectation values by performing a weighted average over the fit models $i$ parameterized by $t_{\rm min}$ and $N$. We use the Akaike Information Criterion (AIC) to determine the relative weights:
 \begin{equation}
       {\rm AIC}_i = \chi_i^2 +2N_{{\rm param},i}-2N_{{\rm data},i}
 \end{equation}
 with $\chi^2$ coming from the fit and $N_{\rm param}$ and $N_{\rm data}$ being the number of parameters and data, respectively. We then perform a weighted average over the fits $i$ with weight
 \begin{equation}
       W_i= \frac{\exp\left({\rm AIC}_i/2\right)}{\sum_i\exp\left({\rm AIC}_i/2\right) }.
 \end{equation}
 We wrap the entire above procedure in a binned jackknife analysis, performing on each sample independent determinations of $M_\pi$, $M_D$, $M_{D^*}$, and the $[6]$ and the $SU(3)$ ground-state mass shifts:

 \begin{eqnarray}
     \Delta E_{[6]_{\rm sca}}&\equiv E_{[6]_{\rm sca}}-\left(M_D+M_\pi\right)\nonumber\\ 
     \Delta E_{[6]_{\rm axv}}&\equiv E_{[6]_{\rm axv}}-\left(M_{D^*}+M_\pi\right)\nonumber \\
    \Delta E_{[\overline{15}]_{\rm sca}}&\equiv E_{[\overline{15}]_{\rm sca}}-\left(M_D+M_\pi\right) \nonumber\\
    \Delta E_{[\overline{15}]_{\rm axv}}&\equiv E_{[\overline{15}]_{\rm axv}}-\left(M_{D^*}+M_\pi\right).\label{eq:deltaEs}
 \end{eqnarray}

\section{Results and interpretation}
\begin{figure}
    \centering
    \includegraphics[width=0.52\linewidth]{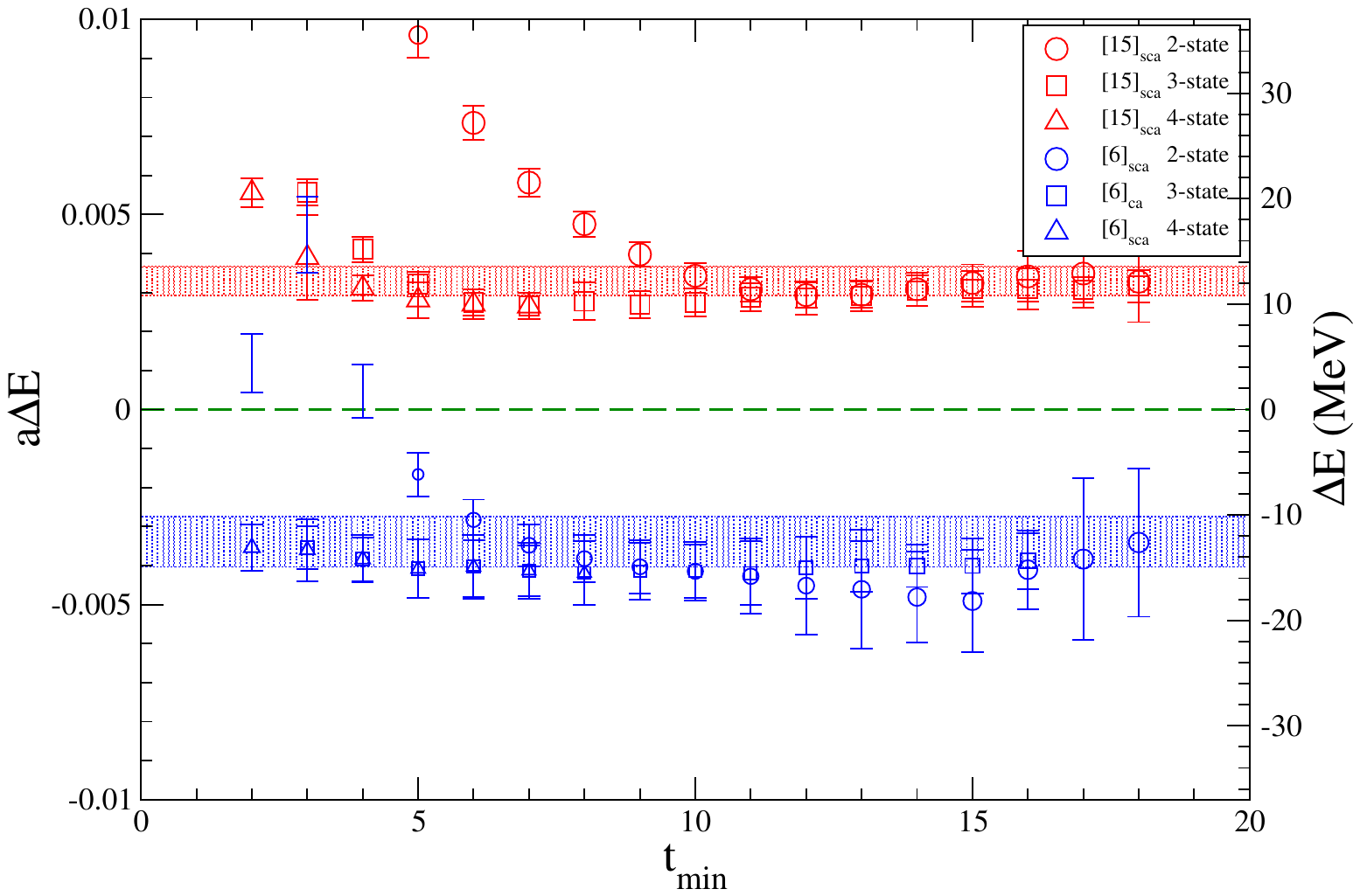}\hspace{-0.3in}
     \includegraphics[width=0.52\linewidth]{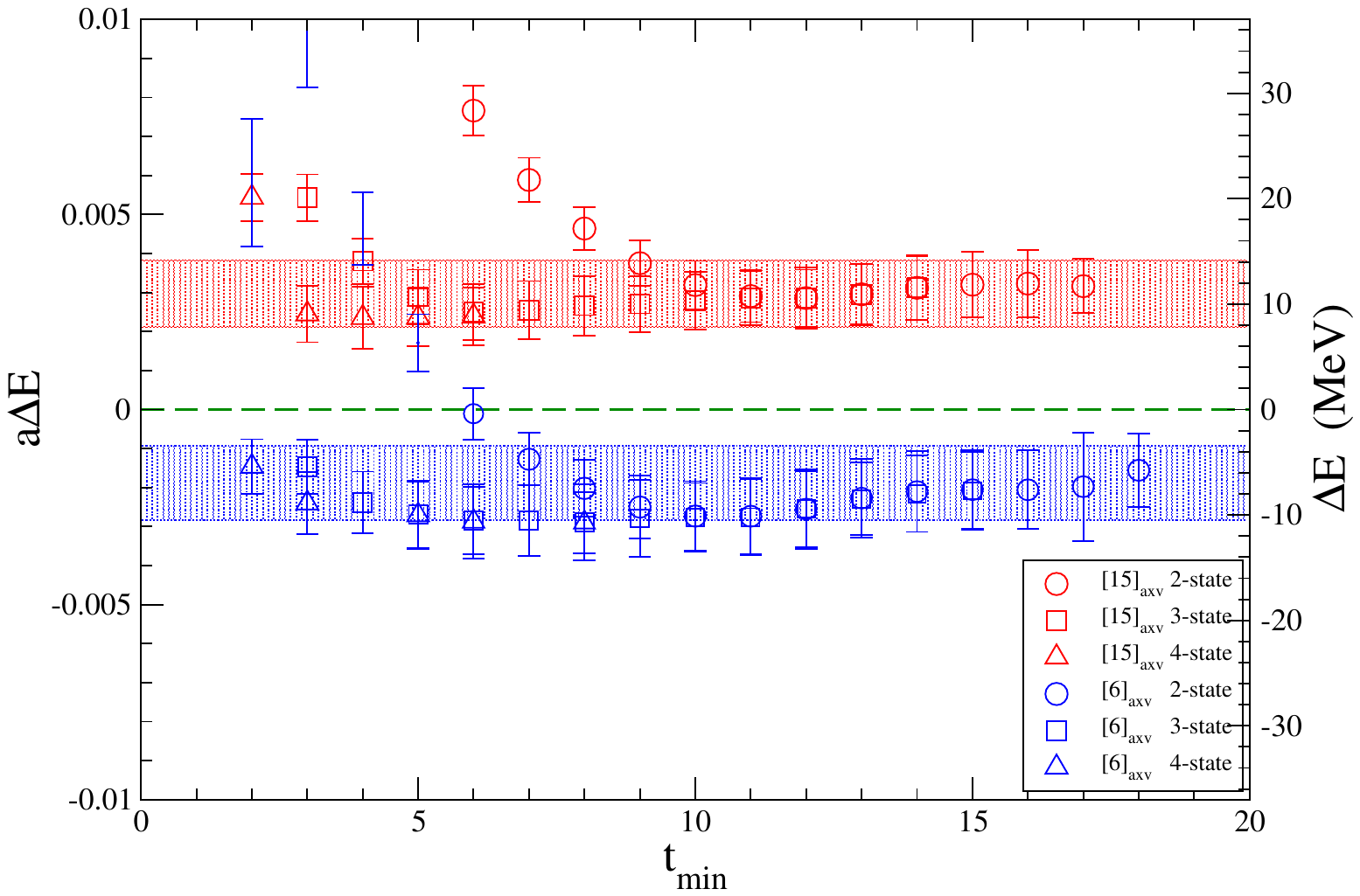}
    \caption{Fit results for the ground-state mass shifts for $[\overline{15}]$ (red) and $[6]$ (blue) for the scalar sector (left) and the axial-vector (right). Circles, squares and triangles represent $N=2,3,4$-state fits, respectively. Shaded bands represent the results of the fit model-averaging procedure and jackknife determination of statistical uncertainty. The green dashed line is the $\pi +D $ or $\pi + D^*$ threshold.}
    \label{fig:results}
\end{figure}

The result of the fitting and analysis procedure is displayed in \autoref{fig:results}. We find:
\begin{eqnarray}
   &\Delta E_{{[6]}_{\rm sca}}&= (-13\pm 2)\text{ MeV}\nonumber\\
        &  \Delta E_{{[6]}_{\rm axv}}&=(-7\pm 4)\text{ MeV}
        \end{eqnarray}
        and
        \begin{eqnarray}
    &\Delta E_{[\overline{15}]_{\rm sca}} &= (12\pm 1)~\text{ MeV} \nonumber \\
    &\Delta E_{[\overline{15}]_{\rm axv}} &= (11\pm 3)~\text{ MeV}.
\end{eqnarray}
It is immediately apparent that we see consistent behavior in both the scalar and axial-vector sectors. 

Particularly significant is that in both cases the $[\overline{15}]$ exhibits repulsive interactions. This is consistent with the expectations from the hadronic molecule model, but  inconsistent with the tetraquark model, which predicts a deeply bound state in the $[\overline{15}]$ in the axial-vector sector, as illustrated in \autoref{fig:guo-hanhart-prediction}. The conclusion is that this lattice calculation eliminates diquark---anti-diquark tetraquarks as a viable model of the positive parity open-charm states. The lattice findings are instead wholly consistent with the predictions of hadronic molecule description.


\section{Acknowledgements}
We are grateful to Giovanni Pederiva for useful discussion on data modeling.
T.L. and S.K. were funded in part by the Deutsche
	Forschungsgemeinschaft (DFG, German Research Foundation) as part of the CRC 1639 NuMeriQS–project no.~511713970. S.K. is further supported by the MKW NRW under funding code NW21-024-A, and by the Helmholtz Association through the AIDAS laboratory.
The authors gratefully acknowledge computing time on the supercomputer JURECA\cite{JURECA} at Forschungszentrum Jülich under Grant ``EXFLASH'', and also 
the Gauss Centre for Supercomputing e.V. (\url{www.gauss-centre.eu}) for funding this project by providing computing time through the John von Neumann Institute for Computing (NIC) on the GCS Supercomputer JUWELS\cite{JUWELS} at Jülich Supercomputing Centre (JSC).
E.B.G. was supported in part by the German Federal Ministry of Education and Research (BMBF) (grant number: 16HPC001) and the Ministry of Culture and Science (MKW) of the state of North-Rhine-Westphalia (grant number: 325-8.03-133340) through funding of the Gauss Centre for Supercomputing (GCS).
F.-K.G. was supported by the National Key R\&D Program of China under Grant No. 2023YFA1606703, by the National Natural Science Foundation of China (NSFC) under Grants No. 12125507 and No. 12047503, and by the Chinese Academy of Sciences (CAS) under Grant No.~YSBR-101. In addition, C.H. thanks the CAS President's International Fellowship Initiative (PIFI) under Grant No.~2025PD0087.

\bibliographystyle{JHEP}
\bibliography{refs}


\end{document}